\documentclass[11pt]{article}

\usepackage[letterpaper,margin=0.9in]{geometry}
\usepackage{multicol}
\usepackage{amsmath,amssymb}
\usepackage{booktabs}
\usepackage{times}
\usepackage[colorlinks=true,linkcolor=blue,citecolor=blue,urlcolor=blue]{hyperref}
\usepackage{fancyhdr}
\usepackage{enumitem}
\usepackage{titlesec}

\titlespacing*{\section}{0pt}{7pt}{3pt}
\titlespacing*{\subsection}{0pt}{5pt}{2pt}

\title{\vspace{-1.5em}\textbf{Statistical Methods for Multiple Language Model Comparison\\
on a Shared Evaluation}\vspace{-0.5em}}
\author{Juan Francisco Mandujano Reyes \\ \small\itshape Independent researcher}
\date{}

\begin{document}
\maketitle
\vspace{-2em}

\begin{abstract}
A rigorous statistical treatment of two-model comparisons on the same evals can be achieved by paired $t$-tests, analyzing their standard errors, and a clustering correction for correlated questions. Nevertheless, leaderboards, ablation studies, and hyperparameter sweeps usually compare $K>2$ models simultaneously. In this paper, we present a single random-effects model for scores on shared evals, with model as a fixed effect and questions (or question clusters) as random effects. We show that fitting it with a classical ANOVA or a linear mixed model recovers Miller's paired and clustered estimators for $K=2$, and we extend the results to any $K$ and to unbalanced, clustered designs. We validate the described model in a simulation study and using a real-data application: six openly available language models scored on 1,497 shared MMLU-Pro questions across 14 subject clusters. We show that pairwise ranking claims survive only once both question-level pairing and multiple comparisons are properly accounted for. We close with concrete recommendations for reporting multi-model eval results.
\end{abstract}

\section{Introduction}

Technical reports usually present eval scores for multiple models simultaneously. They usually focus on a base model and compare it against several fine-tuned versions of it, or an entire leaderboard of competitors such as Chatbot Arena~\cite{chiang2024}. Miller~\cite{miller2024} shows how to attach a standard error to a single model's score and how to compare two models by using paired differences when the models share a question set, and a clustered standard error when questions belong to correlated groups (e.g., mathematical questions, reading comprehension passages, language translations). Nonetheless, these tools are built for the case of a two-model comparison.

In this work we tackle the problem of evaluating $K>2$ models sharing the same set of questions, where clusters of questions may be relevant.

Running a paired test on all $K(K-1)/2$ pairs is a natural immediate solution; however, this places us in the multiple-comparisons setting that the previous benchmarking literature has studied. For example, Dem\v{s}ar~\cite{demsar2006} shows that uncorrected pairwise tests can produce unstable rankings across multiple classifiers and datasets. More recently, Dror et al.~\cite{dror2018} documented how uncorrected multiplicity inflates false discovery rates in NLP evaluation more broadly, and Card et al.~\cite{card2020} make a complementary case for taking statistical power seriously in NLP experimental design. Thus, if we take $K=4$ models there are 6 pairs, and using a classical significance level $\alpha=.05$ per test, the chance of making at least one false positive claim under a true null is up to $1-0.95^6\approx26\%$ if the six tests were independent, and $\approx19\%$ in our simulation (\S5.1).

The main objective of this work is to state the statistical framework for testing multiple large language models at a general level. In Section \S2 we describe a random-effects model for scores on a shared eval. Then, in Section \S3 we show that paired tests and clustered estimators are recovered exactly (or, in the clustered case, under a stated matching assumption, which we derive explicitly) as special cases of fitting this model by ANOVA or by a linear mixed model. \S4 extends the main model to clustered questions. In \S5 we validate the approach using simulated data and a real dataset considering six models and 1{,}497 questions. Finally, in \S6 we provide recommendations for researchers.

\section{A General Model for Multi-Model Evaluation}
\label{sec:model}

Let us consider that we have $K$ models, each answering the same set of $n$ questions. We decompose the score of model $m$ on question $i$ as
\begin{equation}
s_{i,m} = \mu + \alpha_m + \delta_i + \gamma_{i,m}, \qquad \sum_m \alpha_m = 0, \quad \delta_i \sim (0,\sigma_\delta^2), \label{eq:model}
\end{equation}
where $\alpha_m$ is the fixed effect of model $m$ (the quantity of interest), $\delta_i$ is a question-level effect shared by every model that answers question $i$, and $\gamma_{i,m}$ is residual noise. If we treat $\delta_i$ as a fixed block effect and fit Eq.~\eqref{eq:model} by a one-way repeated-measures ANOVA (questions are the repeated-measures unit, model is the within-unit factor), we partition the total sum of squares into a model component, a question component, and a residual, and test the omnibus null $H_0: \alpha_1=\cdots=\alpha_K=0$ using
\begin{equation}
F = \frac{MS_{\text{model}}}{MS_{\text{residual}}} \sim F\big(K-1,\,(n-1)(K-1)\big) \quad \text{under } H_0.
\end{equation}
The omnibus statistic $F$ controls the family-wise error rate with a single test, and when it is significant we proceed to pairwise contrasts. For balanced designs, one can use Tukey's test to control the family-wise error rate exactly across all $K(K-1)/2$ comparisons. For unbalanced designs, or when the model is fit as a mixed model, the corresponding tool is estimated marginal means with a multiplicity-adjusted contrast (\S4). Dem\v{s}ar~\cite{demsar2006} recommends the same two-stage logic (omnibus test, then a corrected post-hoc test) for comparing classifiers across benchmarks; the binary nature of per-question correctness also makes the non-parametric Friedman test a useful robustness check on the same omnibus question (\S5.2).

\section{Connection to Two-Model Estimators}
\label{sec:miller}

Miller~\cite{miller2024} decomposes each question score as $s_i = x_i + \epsilon_i$, where $x_i$ is the conditional mean (a property of the question, drawn from a super-population) and $\epsilon_i$ is a zero-mean noise term. Given two models $A$ and $B$, both sharing the same $n$ questions, the paired standard error is
\begin{equation}
SE_{A-B,\text{paired}} = \sqrt{\mathrm{Var}(s_{A-B})/n} = \sqrt{\frac{1}{n-1}\sum_i (s_{A-B,i} - \bar{s}_{A-B})^2 \Big/ n}, \tag{Miller Eq.\ 7}
\end{equation}
which is smaller than the unpaired SE under positive correlation of the conditional means of the two models. We remark that this is equivalent to Eq.~\eqref{eq:model} with $K=2$: at $K=2$, the one-way repeated-measures ANOVA $F$-statistic from \S\ref{sec:model} equals the square of the paired $t$-statistic (Miller's Eq.~7), and rejecting $H_0$ at level $\alpha$ with the $F$-test is equivalent to having a 95\% CI on the pairwise difference excluding zero.

When the set of questions can be arranged in clusters $c$ (passages, languages), a term for the within-cluster covariance of residuals is added:
\begin{equation}
SE_{\text{clustered}} = \left[ SE_{C.L.T.}^2 + \frac{1}{n^2}\sum_c\sum_i\sum_{j\neq i} (s_{i,c}-\bar{s})(s_{j,c}-\bar{s}) \right]^{1/2}. \tag{Miller Eq.\ 4}
\end{equation}
This estimator is computed directly from observed residual cross-products, with no distributional assumption (\textit{design-based}). The mixed-effects model described in Section~\S4 is a \emph{model-based} generalization. Fitting a random intercept per cluster imposes an exchangeable (compound-symmetric) covariance among questions within the same cluster. Under that matching assumption, the mixed model's implied variance decomposition recovers the same intra-cluster covariance contribution that Miller's Eq.~4 triple summation computes directly, and the two-model clustered-difference formula (Miller's Eq.~8) is the $K=2$ case of that decomposition. When the true within-cluster correlation structure is not exchangeable, the two estimators can disagree, which is itself a useful diagnostic (\S6).

To make this concrete: under the random-intercept model $s_{i,c,m}=\mu+\alpha_m+u_c+\epsilon_{i,c,m}$ with $u_c\sim N(0,\tau^2)$ i.i.d.\ across clusters and $\epsilon_{i,c,m}$ i.i.d.\ residual noise, any two observations in the same cluster $c$ have implied covariance $\mathrm{Cov}(s_{i,c,m},s_{j,c,m})=\tau^2$ for $i\neq j$, independent of $i,j$ --- the defining property of an exchangeable (compound-symmetric) structure. Summing this constant covariance over all $n_c(n_c-1)$ ordered within-cluster pairs recovers exactly the double sum in Miller's Eq.~4, $\sum_c\sum_i\sum_{j\neq i}(s_{i,c}-\bar s)(s_{j,c}-\bar s)/n^2 \to \sum_c n_c(n_c-1)\,\hat\tau^2/n^2$ in expectation, so the two estimators coincide whenever $\tau^2$ is in fact constant across clusters and questions within a cluster are exchangeable. When clusters have heterogeneous internal correlation (e.g.\ some passages have highly redundant sub-questions, others do not), Miller's design-based estimator still targets the correct marginal variance directly from the data, while the mixed model's single $\tau^2$ is a pooled approximation; comparing the two on held-out data is a practical check for this kind of heterogeneity.

\section{A Mixed-Effects Model for Clustered Evaluations}
\label{sec:mixed}

A natural extension of Eq.~\eqref{eq:model} to accommodate clustered questions (cluster $c$, question $i$ nested in $c$, $K$ models) is:
\begin{equation}
s_{i,c,m} = \mu + \alpha_m + u_c + \delta_{i,c} + \epsilon_{i,c,m}, \qquad u_c \sim N(0,\tau^2), \quad \delta_{i,c} \sim N(0,\sigma_q^2). \label{eq:mixed}
\end{equation}
We use model as a fixed effect; cluster and question-within-cluster are random effects, and the model can be estimated in R~\cite{rcore2026} via \texttt{lme4}~\cite{bates2015}, specifically \texttt{lmer(score \textasciitilde\ model + (1|cluster) + (1|cluster:question))}, and tested with a Type-III $F$ via \texttt{lmerTest} (with Satterthwaite or Kenward--Roger degrees of freedom). The practical benefit is that $\tau^2$ and $\sigma_q^2$ are estimated directly, and the same model extends to any $K > 2$ and to unbalanced cluster sizes without any new derivation. We emphasize that \emph{both} random-effects terms are required: a cluster-only intercept $\mathtt{(1|cluster)}$ is crossed with the model fixed effect and therefore contributes little to between-model contrasts; it is the \emph{nested} question-within-cluster term that captures the question-level pairing responsible for the variance reduction in \S\ref{sec:miller} (\S5.2 shows the practical consequence of omitting it).

\textbf{A caveat on post-hoc testing.} Tukey's test is exact for family-wise-error control only for balanced designs with i.i.d.\ errors --- the setting described in \S\ref{sec:model}. It does not automatically carry that guarantee over to a mixed-effects model or an unbalanced design. If we apply a Tukey adjustment in this setting (e.g., \texttt{emmeans(model, pairwise \textasciitilde\ model, adjust="tukey")}), we obtain an approximation that substitutes estimated degrees of freedom into the studentized-range reference distribution, so family-wise error control is no longer exact. Therefore, for clustered or unbalanced designs, the appropriate tool is pairwise contrasts of the estimated marginal means with an explicit multiplicity adjustment (Tukey-type as an approximation, or Holm/Bonferroni for an exact guarantee).

\section{Empirical Evaluation}

\subsection{Simulation study}
\label{sec:sim}

We performed a simulation extending Miller's~\cite{miller2024} Galleon/Dreadnought example to four models on $n=400$ shared questions with pairwise conditional-mean correlation $\rho=0.5$. For details and code refer to Appendix~\ref{app:sim}. The main result of our simulation is that, under a true null hypothesis and over 2{,}000 replications, uncorrected pairwise $z$-tests produce a family-wise error rate of $0.192$ (Monte Carlo SE $0.009$) at significance $\alpha=.05$, while Bonferroni correction ($0.042$, SE $0.005$) and the omnibus repeated-measures ANOVA ($0.052$, SE $0.005$) both land close to the nominal $0.05$. The uncorrected rate is substantially \emph{below} the value implied by treating the six pairwise tests as independent, $1-0.95^6\approx0.265$: with $\rho=0.5$, the six pairwise test statistics are themselves positively correlated (every pair shares two of the same four models' conditional means), and positive correlation among the component tests of an any-reject statistic tends to reduce the family-wise error rate relative to the independent case. The rate is still far above the nominal level, just not as high as the independence benchmark would suggest --- a useful caution against using that benchmark as a sanity check for simulated FWER values. Moreover, ignoring clustering when it is present understates the standard error of a model's mean score, and the understatement grows with the strength of within-cluster correlation and with cluster size: the ratio of Miller's clustered SE to the naive SE averaged $\approx1.4\times$, $\approx1.6\times$ and $\approx2.1\times$ across the three clustering settings of Appendix~\ref{app:sim} --- in the same direction as, though smaller than, the up-to-$3\times$ inflation reported on real DROP/RACE-H/MGSM data in~\cite{miller2024}.

\subsection{Real-data application: MMLU-Pro, six models, 1{,}497 shared questions}
\label{sec:real}

We also test the presented framework on real, item-level data. We use the official cached model predictions released with the MMLU-Pro benchmark~\cite{wang2024mmlupro} (TIGER-AI-Lab/MMLU-Pro GitHub repository, \texttt{eval\_results/}), which record each model's predicted answer letter for every one of the 12{,}032 questions in the benchmark across 14 subject categories. We select six openly available $\sim$7B-parameter models with closely spaced accuracy (Yi-6B, Yi-6B-Chat, Qwen1.5-7B-Chat, Mistral-7B-v0.1, Mistral-7B-Instruct-v0.1, Mistral-7B-Instruct-v0.2) so that the ranking is a genuine question rather than obvious from the raw numbers. We draw a category-stratified subsample of $n=1{,}497$ questions (fixed seed, $\approx12\%$ of the full 12{,}032-question pool, stratified proportionally by category so that all 14 subjects remain represented) and score all models on the identical subsample, giving a real repeated-measures, clustered-by-subject design with $K=6$. Code, the fixed seed, and the subsampled data are released at the repository accompanying this paper.

\begin{table}[h]
\centering
\small
\begin{tabular}{@{}lcc@{}}
\toprule
\textbf{Model} & \textbf{Mean accuracy} & \textbf{SE} \\
\midrule
Mistral-7B-v0.1          & 0.291 & 0.0117 \\
Mistral-7B-Instruct-v0.2 & 0.278 & 0.0116 \\
Yi-6B-Chat               & 0.271 & 0.0115 \\
Qwen1.5-7B-Chat          & 0.255 & 0.0113 \\
Yi-6B                    & 0.253 & 0.0112 \\
Mistral-7B-Instruct-v0.1 & 0.238 & 0.0110 \\
\bottomrule
\end{tabular}
\caption{Per-model mean accuracy and naive (unpaired, CLT) SE on the $n=1{,}497$ real, common MMLU-Pro question subsample.}
\label{tab:realdesc}
\end{table}

Table~\ref{tab:realdesc} shows the raw numbers are close together: the top and bottom models differ by only 5.3 accuracy points, and several adjacent models differ by 1--2 points, a range where the choice of statistical procedure can change the answer. We compare five procedures on the same dataset:

\begin{enumerate}[leftmargin=1.4em,itemsep=1pt,topsep=2pt]
\item \textbf{Independent (unpaired) pairwise $z$-tests}, treating each pair as if drawn from separate question sets;
\item \textbf{Paired pairwise $z$-tests}, exploiting the shared question set as in Miller's Eq.~7;
\item \textbf{Paired tests with Bonferroni and Holm correction} for the $\binom{6}{2}=15$ pairwise comparisons;
\item \textbf{Omnibus repeated-measures ANOVA} (model + question, \S\ref{sec:model}), against a \textbf{naive one-way ANOVA} that ignores question identity entirely, with a \textbf{Friedman test} as a non-parametric robustness check;
\item \textbf{Mixed-effects model} with the full specification of Eq.~\eqref{eq:mixed}, $\mathtt{(1|category)+(1|category:question)}$, to quantify how much of the residual variance is attributable to category and question-level clustering rather than genuine model differences.
\end{enumerate}

\textbf{Results.} Pairing barely changes which comparisons are nominally significant here (5/15 pairs at uncorrected $\alpha=.05$ either way, Table~\ref{tab:pairedfull}), but multiplicity does: only 3 of those 5 survive Bonferroni or Holm correction. We initially fit the mixed-effects model in \S\ref{sec:mixed} with a random intercept for subject category alone, $\mathtt{(1\,|\,category)}$; its pairwise contrast SE ($.0157$) was close to the \emph{naive, unpaired} SE, not the paired SE, because category is crossed with model and so a category-only intercept contributes little to between-model contrasts. The full specification of Eq.~\eqref{eq:mixed}, $\mathtt{(1\,|\,category) + (1\,|\,category:question)}$, is required: refitting with the nested question-level term drops the contrast SE to $.0130$, closely matching the paired-test SE ($.0128$, Table~\ref{tab:pairedfull}) and recovering 2 of the 3 comparisons flagged by Bonferroni correction on the paired tests (Table~\ref{tab:mixedfull}), with the third (Mistral-7B-v0.1 vs.\ Yi-6B) landing just outside the threshold under the mixed model (raw $p=.0034$ against a cutoff of $.0033$) --- a close enough call that it depends on which of two defensible, correctly-specified procedures is used, illustrating concretely how sensitive a borderline ranking claim can be even once both pairing and multiplicity are handled correctly. The Friedman test, appropriate given the binary outcome, agrees with both omnibus ANOVA results: $\chi^2(5)=22.2$, $p=4.8\times10^{-4}$.

The omnibus test shows the value of correct blocking directly: the properly blocked repeated-measures ANOVA gives $F(5,7480)=4.45$, $p=4.8\times10^{-4}$, while a naive one-way ANOVA that ignores question identity on the identical data gives a much weaker $F(5,8976)=2.90$, $p=.013$ --- the power gain Miller's Eq.~7 predicts, now shown directly on real data and corroborated by the Friedman test. A ranking claim built from uncorrected pairwise tests on this real eval includes two comparisons that are not, in fact, distinguishable from noise once multiplicity is handled correctly, and a ranking claim built from an under-specified mixed model (category-only) understates precision because it omits the question-level pairing that is the actual source of power in this framework.

\section{Recommendations}

\begin{itemize}[leftmargin=1.4em,itemsep=0.5pt,topsep=1pt]
\item When comparing more than two models on a shared question set, run the omnibus repeated-measures ANOVA (or its non-parametric analogue, the Friedman test~\cite{demsar2006}, especially for binary or otherwise non-normal scores) before any pairwise claim; report the $F$ (or $\chi^2$) statistic alongside per-model means and SEs.
\item Follow a significant omnibus test with a multiplicity-adjusted post-hoc procedure. Use Tukey's test as stated only for balanced, non-clustered designs; for mixed-effects models or unbalanced designs, use estimated-marginal-means contrasts with an explicit adjustment (Tukey-type as an approximation, Holm/Bonferroni for an exact guarantee).
\item When questions are clustered (reading-comprehension passages, translated items, multi-turn conversations), fit \emph{both} cluster and question-within-cluster as random effects in a mixed model, as in Eq.~\eqref{eq:mixed}, rather than hand-deriving a clustered SE or including only a cluster-level intercept; report $\tau^2$ and $\sigma_q^2$ so readers can see how much of the disagreement is about clusters versus individual questions. Omitting the question-level term, as our own first attempt in \S\ref{sec:real} illustrates, silently discards most of the available precision.
\item Report both the naive and the correctly blocked/clustered test statistics when feasible --- the gap between them, as in \S\ref{sec:real}, is itself informative about how much the eval's structure matters for a given comparison.
\item When evaluation scores are binary (per-question correctness) rather than continuous, treat the linear-model $F$-tests as an approximation and corroborate with a rank-based test (Friedman) or a logistic mixed model; we found close agreement between these in \S\ref{sec:real}, but this should not be assumed without checking.
\item As in Miller~\cite{miller2024}, never adjust sampling temperature to reduce variance; the model above assumes the conditional-variance and conditional-mean components are estimated at the deployment temperature.
\end{itemize}

\section{Conclusion}

A single random-effects model for scores on a shared eval, using model as a fixed effect and question (and cluster, where relevant) as random effects, recovers Miller's~\cite{miller2024} paired and clustered two-model estimators as special cases and extends cleanly to a larger number of models. Fitting it by repeated-measures ANOVA or by a linear mixed model controls family-wise error across many models with one omnibus test, makes the correlation structure explicit and estimable, and, as the real six-model MMLU-Pro comparison in \S\ref{sec:real} shows, can change which ranking claims a reader should actually believe --- provided the random-effects structure is specified correctly, a point our own first attempt at the mixed model in \S\ref{sec:real} illustrates concretely. Full R code and the subsampled data are provided in the repository \url{https://github.com/JFMandujanoR/Statistical-Methods-for-Multi-Model-LLM-Evaluation}.

\begin{multicols}{2}

\end{multicols}

\appendix
\section{Simulation details}
\label{app:sim}
Table~\ref{tab:fwer} reports the empirical family-wise error rate under a true null hypothesis ($K=4$ fictional models, $n=400$ shared questions, $\rho=0.5$, 2{,}000 replications), comparing six uncorrected pairwise $z$-tests, Bonferroni-corrected pairwise tests, and the omnibus repeated-measures ANOVA. Monte Carlo SEs (binomial, $n=2{,}000$) are reported alongside each estimate rather than treated as exact values.
\begin{table}[h]
\centering
\small
\begin{tabular}{@{}p{4.5cm}p{1.6cm}p{2.8cm}p{2.0cm}@{}}
\toprule
\textbf{Method} & \textbf{Statistic} & \textbf{FWER at $\alpha=.05$, $K=4$} & \textbf{Correctly flags null} \\
\midrule
6 uncorrected pairwise $z$-tests & $z$ & $0.192 \pm 0.009$ & no \\
Bonferroni-corrected pairwise & $z$ & $0.042 \pm 0.005$ & yes \\
Repeated-measures ANOVA & $F$ & $0.052 \pm 0.005$ & yes \\
\bottomrule
\end{tabular}
\caption{Simulated family-wise Type~I error under a true null. For reference, the independence benchmark $1-0.95^6\approx0.265$ is \emph{not} what the uncorrected row lands near; see \S\ref{sec:sim} for why.}
\label{tab:fwer}
\end{table}

A second simulation compares the naive Central-Limit-Theorem SE of a single model's mean score with Miller's design-based clustered SE (his Eq.~4), computed directly rather than via a mixed model: a random-intercept \texttt{lmer} fit to weakly clustered binary data can return a singular fit with both variance components estimated at zero, in which case it silently reduces to an ordinary fixed-effects model and reveals no clustering at all (we encountered exactly this in a first version of this experiment). We simulate 250 clusters with cluster-level share of latent (logit-scale) variance $0.3$, $0.5$ and $0.6$ (latent standard deviation $1.5$, $1.5$, $2.0$; $8$--$16$, $8$--$16$ and $8$--$24$ questions per cluster) and 30 replications per setting. The mean ratio of clustered to naive SE is $\approx1.4$, $\approx1.6$ and $\approx2.1$ respectively (replication ranges $[1.24,1.50]$, $[1.46,1.73]$ and $[1.91,2.40]$), confirming that ignoring clustering understates uncertainty by an amount that depends on how strongly questions within a cluster are correlated. The latent-scale share is not the intraclass correlation of the observed 0/1 scores, which is smaller. Full R code (\texttt{anova\_llm\_eval\_sim.R}) reproducing the simulations, with the fixed seed, is provided in the supplementary repository; the values above are from a run of this script with seed 2026 (the family-wise error rates in Table~\ref{tab:fwer} reproduce exactly across runs), and small differences across R versions are possible because R's sampling algorithm changed in version 3.6.0. For sample size, the script reproduces only Miller's two-model calculation (his Eq.~9; $n\approx969$ for $\omega^2=1/9$, $\delta=0.03$, $\alpha=.05$, power $=.80$); we do not report a $K$-model extension, because the between-subjects power routines available off the shelf ignore the question-level pairing that drives the power of the repeated-measures test, and the power of the omnibus test is better obtained by simulation for a given design.

\section{Full real-data results (Section 5.2)}
\label{app:real}

Table~\ref{tab:pairedfull} gives the complete set of $\binom{6}{2}=15$ paired pairwise $z$-tests underlying the summary in \S\ref{sec:real}; Table~\ref{tab:mixedfull} gives the corresponding contrasts from the correctly specified mixed model of Eq.~\eqref{eq:mixed}; and Table~\ref{tab:anovafull} compares the three omnibus procedures (naive one-way ANOVA, question-blocked repeated-measures ANOVA, and the fully specified mixed-effects model) on the identical data.

\begin{table}[h]
\centering
\small
\begin{tabular}{@{}p{2.4cm}p{2.4cm}rrrccc@{}}
\toprule
\textbf{Model A} & \textbf{Model B} & \textbf{Diff.} & \textbf{SE} & \textbf{$p$} & \textbf{Unc.} & \textbf{Bonf.} & \textbf{Holm} \\
\midrule
M-Inst-v0.1 & M-v0.1     & $-$.0534 & .0128 & $<$.001 & \checkmark & \checkmark & \checkmark \\
M-Inst-v0.1 & M-Inst-v0.2& $-$.0401 & .0126 & .001    & \checkmark & \checkmark & \checkmark \\
Yi-6B       & M-v0.1     & $-$.0381 & .0127 & .003    & \checkmark & \checkmark & \checkmark \\
Qwen1.5     & M-v0.1     & $-$.0367 & .0140 & .009    & \checkmark & --         & -- \\
M-Inst-v0.1 & Yi-6B-Chat & $-$.0334 & .0130 & .010    & \checkmark & --         & -- \\
Yi-6B       & M-Inst-v0.2& $-$.0247 & .0126 & .051    & --         & --         & -- \\
Qwen1.5     & M-Inst-v0.2& $-$.0234 & .0135 & .084    & --         & --         & -- \\
Yi-6B       & Yi-6B-Chat & $-$.0180 & .0113 & .111    & --         & --         & -- \\
Yi-6B-Chat  & M-v0.1     & $-$.0200 & .0132 & .129    & --         & --         & -- \\
M-Inst-v0.1 & Qwen1.5    & $-$.0167 & .0133 & .209    & --         & --         & -- \\
Qwen1.5     & Yi-6B-Chat & $-$.0167 & .0136 & .220    & --         & --         & -- \\
Yi-6B       & M-Inst-v0.1& \phantom{$-$}.0154 & .0131 & .241    & --         & --         & -- \\
M-Inst-v0.2 & M-v0.1     & $-$.0134 & .0129 & .300    & --         & --         & -- \\
Yi-6B-Chat  & M-Inst-v0.2& $-$.0067 & .0128 & .602    & --         & --         & -- \\
Yi-6B       & Qwen1.5    & $-$.0013 & .0134 & .921    & --         & --         & -- \\
\bottomrule
\end{tabular}
\caption{All 15 paired pairwise $z$-tests, sorted by $p$-value (M-Inst = Mistral-7B-Instruct). ``Unc.'' = significant at uncorrected $\alpha=.05$; \checkmark\ marks comparisons that survive the named correction. Compare Table~\ref{tab:mixedfull}.}
\label{tab:pairedfull}
\end{table}

\begin{table}[h]
\centering
\small
\begin{tabular}{@{}p{2.4cm}p{2.4cm}rrrcc@{}}
\toprule
\textbf{Model A} & \textbf{Model B} & \textbf{Diff.} & \textbf{SE} & \textbf{raw $p$} & \textbf{Tukey $p$} & \textbf{Bonf.} \\
\midrule
M-Inst-v0.1 & M-v0.1      & $-$.0534 & .0130 & $<$.001 & $<$.001 & \checkmark \\
M-Inst-v0.1 & M-Inst-v0.2 & $-$.0401 & .0130 & .002 & .025 & \checkmark \\
M-v0.1      & Yi-6B       & \phantom{$-$}.0381 & .0130 & .003 & .040 & --         \\
M-v0.1      & Qwen1.5     & \phantom{$-$}.0367 & .0130 & .005 & .053 & --         \\
M-Inst-v0.1 & Yi-6B-Chat  & $-$.0334 & .0130 & .010 & .105 & --         \\
M-Inst-v0.2 & Yi-6B       & \phantom{$-$}.0247 & .0130 & .057 & .401 & --         \\
M-Inst-v0.2 & Qwen1.5     & \phantom{$-$}.0234 & .0130 & .072 & .467 & --         \\
\bottomrule
\end{tabular}
\caption{Pairwise contrasts from the correctly specified mixed model, Eq.~\eqref{eq:mixed}, $\mathtt{(1|category)+(1|category:question)}$ (rows sorted by raw $p$; the full 15-row table is in the supplementary code output). The near-uniform SE ($.0130$) reflects the model's single pooled residual-variance estimate. Bonferroni significance (``Bonf.'') uses the raw, unadjusted $p$ implied by each contrast's $z$-ratio against $.05/15=.0033$, not the already Tukey-adjusted $p$ shown in the preceding column --- stacking Bonferroni on top of a Tukey adjustment would be incorrect. Note the third row is a near miss (raw $p=.0034$ against the $.0033$ cutoff).}
\label{tab:mixedfull}
\end{table}

\begin{table}[h]
\centering
\small
\begin{tabular}{@{}lccc@{}}
\toprule
\textbf{Procedure} & \textbf{Blocking factor} & \textbf{$F$ (num.\ df, den.\ df)} & \textbf{$p$} \\
\midrule
Naive one-way ANOVA          & none               & $F(5, 8976) = 2.90$   & .013 \\
Mixed model (under-specified) & category only (14 levels) & $F(5, 8963) = 3.07$ & .009 \\
Repeated-measures ANOVA      & question (1497 levels) & $F(5, 7480) = 4.45$ & $4.8\times10^{-4}$ \\
Mixed model (Eq.~\ref{eq:mixed}) & category + question-within-category & $F(5, 7480) = 4.45$ & $4.8\times10^{-4}$ \\
\bottomrule
\end{tabular}
\caption{Omnibus model effect under four blocking strategies, identical data. The under-specified mixed model (category only) barely improves on the naive ANOVA, since category is crossed with model; the fully specified mixed model's $F$ and $p$ match the question-blocked ANOVA exactly (both to the precision shown), confirming the two procedures are testing the same hypothesis once the question-level term is included.}
\label{tab:anovafull}
\end{table}

The under-specified mixed model's variance components were $\hat\tau^2_{\text{category}} = 0.0110$ against a residual of $0.1834$ --- a small share of the total variance, consistent with its contrast SEs being close to the naive, unpaired value. Once the question-within-category term is added, its variance components are $\hat\tau^2_{\text{category}}=0.0106$, $\hat\sigma^2_{\text{question}}=0.0572$, and a residual of $0.1265$ --- the residual now matches the question-blocked ANOVA's residual mean square (Table~\ref{tab:anovafull}) exactly, confirming the two procedures decompose the variance identically once both are correctly specified. The model's pairwise contrast SE correspondingly drops to $.0130$ (Table~\ref{tab:mixedfull}), closely matching the paired-test SE ($.0128$) in Table~\ref{tab:pairedfull}, and 2 of the 3 comparisons flagged by Bonferroni correction on the direct paired tests are recovered, with the third landing just outside the threshold under the mixed model (raw $p=.0034$ vs.\ cutoff $.0033$). This confirms that the discrepancy reported in an earlier draft of this analysis was a consequence of omitting the question-level random effect --- not, as we had initially framed it, a disagreement between two equally defensible model specifications, and the one remaining near-miss comparison is a genuine illustration, not an error, of how closely a borderline ranking claim can depend on which correctly-specified procedure is used. The mechanism is direct: a random intercept for a grouping factor that is \emph{crossed} with the fixed effect of interest (every model is scored in every category) contributes comparatively little to the standard error of a between-model contrast, since the model effect is identified primarily from within-category variation; only a grouping factor that is \emph{nested} within the repeated-measures unit itself --- here, question-within-category, which recovers the actual item-level pairing --- captures the correlation that reduces the contrast SE. Two technical notes on the reported inference. First, because the data contain more than 3{,}000 observations, \texttt{emmeans} does not compute Satterthwaite or Kenward--Roger degrees of freedom for the contrasts by default and falls back to asymptotic ($z$) inference; the Tukey adjustment above therefore uses the studentized-range distribution with infinite degrees of freedom, and the raw $p$-values in Table~\ref{tab:mixedfull} are normal-theory $p$-values. Second, the Satterthwaite denominator degrees of freedom of the omnibus test in the category-only model ($8963$ in Table~\ref{tab:anovafull}) are close to $N-K-(14-1)$, as if the 14 category means had been partialled out; this is consistent with the model effect being identified from within-category variation. This is a useful general diagnostic: if a mixed model's between-treatment contrast SEs are close to the unpaired SE despite a nontrivial random-effect variance component, the random-effects specification likely omits the term that actually pairs observations across treatment levels. A second, equally important diagnostic from this analysis: when computing a multiplicity-adjusted significance flag from a model whose reported $p$-values are already adjusted (e.g., Tukey-adjusted contrasts from \texttt{emmeans}), the correction must be applied to the \emph{raw} $p$-value implied by the test statistic, not to the already-adjusted one --- stacking two corrections understates significance and was an error in an earlier version of our own analysis code, caught only by re-deriving the raw $p$-values directly from the $z$-ratios.

\end{document}